\documentclass[a4paper,10.5pt,twocolumn,twoside]{article}

\usepackage{graphicx}
\usepackage{color}
\usepackage[a4paper, left=1.4cm, right=1cm, top=1.7cm, bottom=1cm]{geometry}
\usepackage[super,comma,sort&compress,square]{natbib}
\usepackage{amsmath}
\usepackage{float}
\usepackage{latexsym}
\usepackage{amssymb}

\usepackage{fancyhdr}
\usepackage[utf8x]{inputenc}
\usepackage[T1]{fontenc}
\begin{document}
\pagestyle{fancy}
\def\headrulewidth{0.5pt}
\def\footrulewidth{0pt}
\lhead{ACS Applied Materials \& Interfaces 8 (2016), 28159 -- 28165} 
\chead{}
\rhead{DOI: 10.1021/acsami.6b08532}

\lfoot{} 
\cfoot{}
\rfoot{}

\twocolumn[
  \begin{@twocolumnfalse}
  {\huge \bf The influence of superparamagnetism on exchange anisotropy at CoO/$\lbrack$Co/Pd$\rbrack$ interfaces}

  \hspace{1.1cm}
  \parbox{.87\textwidth}{
    \vspace{4ex}
%    \begin{center}
    \Large \textsf{Marcin Perzanowski,$^{1}$ Marta Marszalek,$^{1}$ Arkadiusz Zarzycki,$^{1}$ Michal Krupinski,$^{1}$ Andrzej Dziedzic,$^{2}$ Yevhen Zabila$^{1}$}
    \vspace{1ex} \\
    \normalsize $^{1}$ Institute of Nuclear Physics Polish Academy of Sciences, Deparment of Materials Science, Radzikowskiego 152, 31-342 Krakow, Poland  \\
    $^{2}$ University of Rzeszow, Center of Innovation \& Knowledge Transfer, Pigonia 1, 35-310 Rzeszow, Poland
%    \end{center}
    \vspace{1ex} \\
    \normalsize \text{email: Marcin.Perzanowski@ifj.edu.pl}

    \vspace{2ex} 
    \noindent
     \textbf{Abstract}: Magnetic systems exhibiting exchange bias effect are being considered as materials for applications in data storage devices, sensors and biomedicine. 
      As the size of new mag- netic devices is being continuously reduced, the influence of thermally induced instabilities in magnetic order has to be taken into account during their fabrication process. 
      In this study we show the influence of superparamagnetism on magnetic properties of exchange-biased $\lbrack$CoO/Co/Pd$\rbrack_{10}$ multilayer. 
      We find that the process of progressive thermal blocking of the superparamagnetic clusters causes an unusually fast rise of the exchange anisotropy field and coercivity, and promotes the easy axis switching to out-of-plane direction.
     
     \vspace{2ex}
     DOI: 10.1021/acsami.6b08532
     
     \vspace{2ex}
     
    \vspace{3ex}
  }
  \end{@twocolumnfalse}
]

\section{Introduction}

Exchange bias is a magnetic effect which usually appears at an interface between ferromagnetic (FM) and antiferromagnetic (AFM) materials.\cite{Kiw01} 
The magnetic hysteresis loop of the exchange-biased system is centered around a~non-zero magnetic field and the loop shift along the external magnetic field axis is called exchange anisotropy field $H_{\mathrm{ex}}$. 
This effect is driven by the exchange anisotropy occurring when the FM/AFM system is field-cooled through the AFM N\'eel temperature $T_{\mathrm{N}}$ lower than Curie temperature $T_{\mathrm{C}}$ of the FM. 
Exchange bias vanishes above the blocking temperature $T_{\mathrm{b}}$ which is typically lower than bulk AFM N\'eel temperature due to finite-size effects at the interface.\cite{Nog99,Nog05} 
Due to the asymmetry in the magnetic reversal process magnetic systems exhibiting exchange bias are being considered as materials for applications in magnetic sensors, storage devices and memories,\cite{Zho04,Sta00,Pol14} as well as in biomedicine as drug carriers.\cite{Kle07,Iss13}

The exchange anisotropy field $H_{\mathrm{ex}}$ is inversely proportional to the thickness of the FM layer~\cite{Nog99} indicating that the exchange bias effect has an interfacial origin. 
This property opens the way for manipulating the exchange anisotropy and the magnetization reorientation mechanism.
A lot of experimental work~\cite{Nog99,Nog05,Ber99} has been carried out on the AFM/FM multilayers for systems with FM layer thickness in nm range.
However, multilayers having lower FM thickness are still less well studied.
Decreasing the size of a FM leads to the superparamagnetism when the magnetic anisotropy is comparable to the thermal fluctuation energy.
In such case the orientation of magnetic moments is thermally unstable.
This makes the system inappropriate for industrial and biomedical application.
In this study we want to address the question of the relationship between progressive blocking of the superparamagnetic particles and the exchange anisotropy on the example of the exchange-biased multilayer with FM volume below the limit for continuous layer formation.
In this case superparamagnetic clusters are expected to occur, allowing the investigation of their magnetic coupling to the AFM material resulting in exchange anisotropy energy.

Our research was carried out on $\lbrack$CoO/Co/Pd$\rbrack_{10}$ polycrystalline multilayer. 
In this structure the AFM/FM CoO/Co interface is a model system for the exchange bias phenomena investigations \cite{Men14,Dob12,Gru00}, and is responsible for introducing the exchange anisotropy. 
The perpendicular magnetic anisotropy, important for applications of the exchange-biased systems in data storage and memories, is implemented in the Co-Pd interface in multilayers or alloys.\cite{Car85,Car03}
Our study shows that lowering the FM volume below the superparamagnetic limit results in large exchange anisotropy field and in its unusual dependence on temperature. 
Moreover, the progressive blocking of the particles also affects the coercivity of the system causing its fast rise with decreasing temperature, and causes the easy axis switching into the out-of-plane direction. 

\section{Experimental section}

The investigated system was fabricated in ultra-high vacuum by thermal evaporation at room temperature under the pressure of $10^{-7}$~Pa. 
Before the multilayer deposition the Si(100) single crystal substrate was covered by a~$5$~nm thick Pd buffer layer. 
The CoO layer was made from $0.5$~nm thick Co layer by exposing it for $10$~minutes to a~pure oxygen atmosphere at the pressure of $3 \times 10^{2}$~Pa. 
Next, the oxide layer was covered by $0.3$~nm of Co and $0.9$~nm of Pd. 
After 10 repetitions of the CoO/Co/Pd trilayer $2$~nm of Pd was deposited as a~capping layer. 

The X-ray reflectivity (XRR) studies were carried out using X'Pert Pro PANalytical diffractometer equipped with Cu X-ray tube operated at 40 kV and 30 mA. 
The transmission electron microscopy (TEM) studies were performed with FEI Tecnai Osiris device. 
The primary electron beam was accelerated using a~voltage of 200~kV. 

Magnetic measurements were performed using Quantum Design MPMS XL SQUID magnetometer. 
The zero field cooling (ZFC) and field cooling (FC) magnetization curves were obtained using a~standard protocol. 
After demagnetization at 300~K the system was cooled down to 5~K without magnetic field. 
Then, the external magnetic field of 500~Oe was applied and the ZFC curve was recorded during heating up to 300~K. 
The FC curve was measured during cooling from 300~K down to 5~K in the same external magnetic field. 
The time-dependent magnetic relaxation curves were obtained at 10~K after switching off the magnetic field from $+50$~kOe. 
The hysteresis loops were measured at different temperatures after cooling the system in the magnetic field of $+50$~kOe. 
The data were corrected for diamagnetic background from sample holder.

\section{Results and discussion}

\begin{figure*}[t]
\centering
\includegraphics[width=0.8\textwidth]{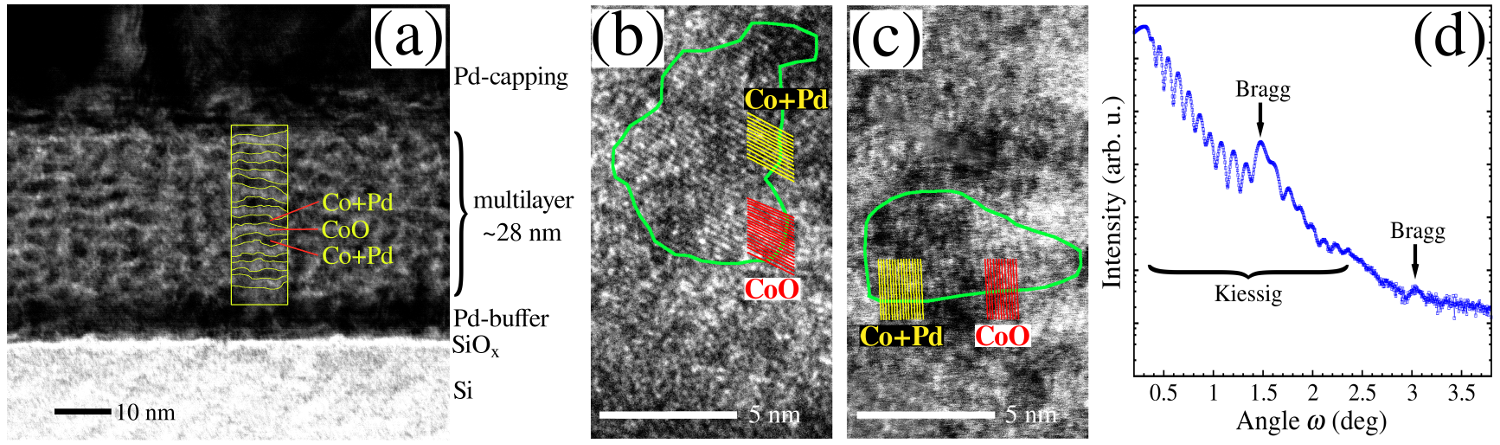}
\caption{(a) TEM cross-section of the $\lbrack$CoO/Co/Pd$\rbrack_{10}$ system. A~representative region where the periodic stacking of the constituent layers is present is indicated in yellow. (b) and (c) Representative multilayer regions (green lines) where the ordered Co-Pd clusters are coupled with the CoO grains. The red and yellow lines mark the atomic planes within CoO and Co-Pd. 
(d) XRR measurement of the $\lbrack$CoO/Co/Pd$\rbrack_{10}$ system. The Bragg reflections and Kiessig fringes are indicated.}
\label{Fig_1}
\end{figure*}

The constitution of the multilayer was investigated with transmission electron microscopy (TEM) and X-ray reflectivity (XRR), and the results are shown in~\ref{Fig_1}. 

The TEM cross-section of the system is presented in~\ref{Fig_1}a.  
The interfaces are highly jagged and the layers are not continuous, however; the alternating stacking of the interfaces along the growth direction is preserved.
The total multilayer thickness is approximately 28~nm. 
Due to the small thickness of the deposited Co it is impossible to set apart this material from Pd and CoO layers, and therefore the multilayer structure will be considered as $[$CoO/Co-Pd$]_{10}$ with Co atoms intermixed with Pd layers. 
The size of the CoO regions ranges from 1~nm to 1.7~nm while the thickness of the Co-Pd regions is between 0.7~nm and 1.3~nm. 
Representative multilayer regions, inwhich the CoO grains are in the vicinity of the Co-Pd clusters, are presented in~\ref{Fig_1}b~and~1c. 
The size of the CoO/Co-Pd regions with periodic stacking of the crystallographic planes is in a~range of a~few nm and differs from cluster to cluster. 

The XRR measuremet is shown in~\ref{Fig_1}d. 
The Kiessig fringes are observed up to 2.5$^{\circ}$ and the angular distances between two neighboring maxima correspond to the total system thickness of approximately 35~nm. 
Taking into account the thicknesses of the buffer and capping Pd layers the thickness of the $\lbrack$CoO/Co/Pd$\rbrack_{10}$ multilayer can be estimated as 28~nm which remains in agreement with the TEM data.   
The presence of two Bragg reflections indicates a~periodic change of the electron density along the growth direction with the period thickness of 2.8~nm. 
It was found by Rafaja et al.~\cite{Raf02} that even in case of non-continuous layers the Bragg reflection should be observed in the XRR curve, which is consistent with our results. 
Due to the jaggedness of the interfaces the XRR data cannot be well-fitted with a~multilayer structure model which precludes a~precise determination of the densities of the constituent materials.

\begin{figure*}[t]
\centering
\includegraphics[width=0.5\textwidth]{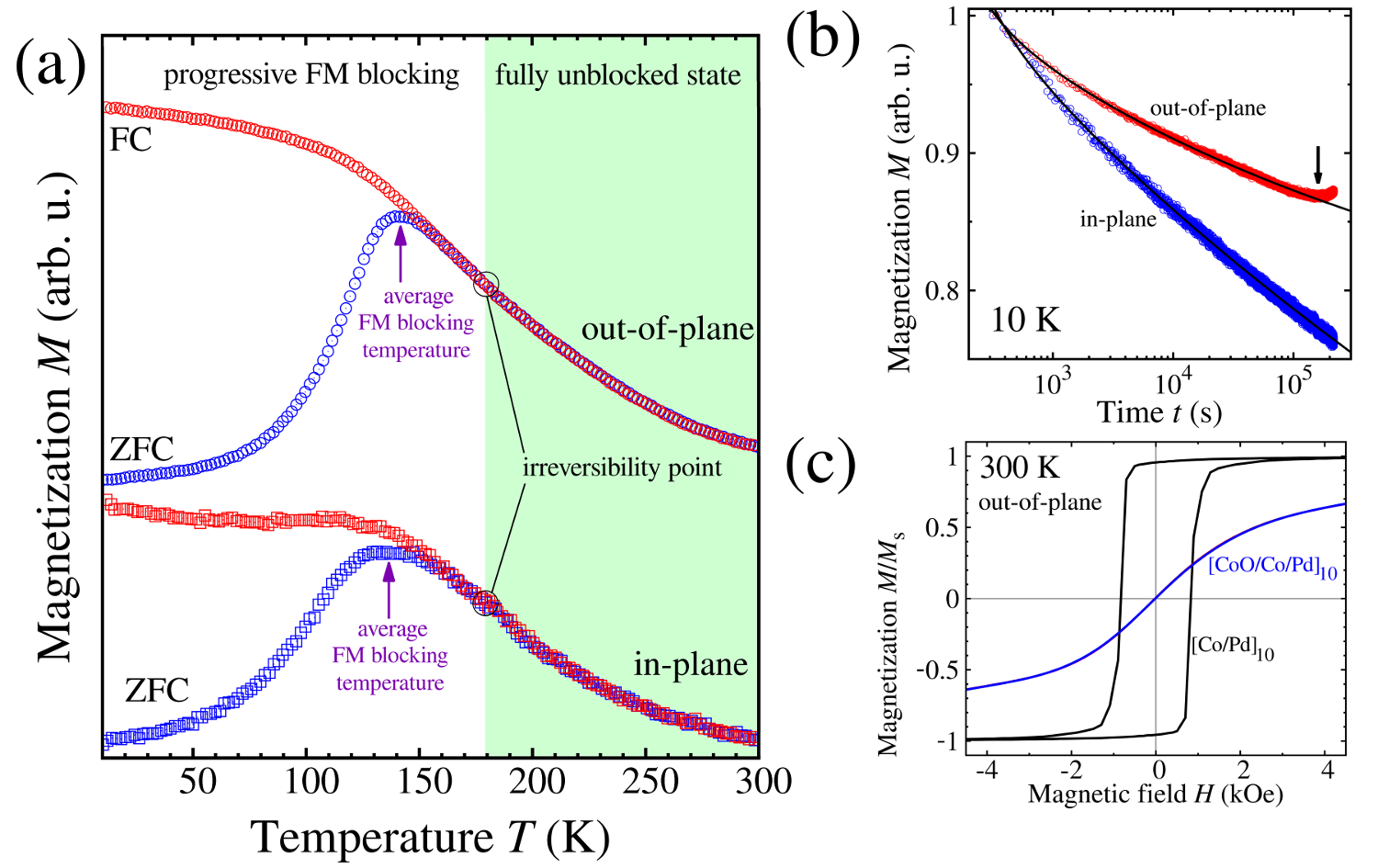}
\caption{(a) Zero field cooling (ZFC) and field cooling (FC) measurements for $\lbrack$CoO/Co/Pd$\rbrack_{10}$ multilayer for out-of-plane and in-plane geometries. The curves were obtained with external magnetic field of $500$~Oe. The measurements were shifted vertically for clarity. (b) Magnetic relaxation curves measured at 10~K for in-plane and out-of-plane geometries. Solid lines are fits (see text). The arrow indicates the minimum of the magnetization. (c) Hysteresis loops for [Co/Pd]$_{10}$ (black line) and [CoO/Co/Pd]$_{10}$ (blue line) multilayers measured at 300~K in out-of-plane geometry.}
\label{Fig_2}
\end{figure*}

The field cooling (FC) and zero field cooling (ZFC) measurements, presented in~\ref{Fig_2}a, were carried out for in-plane (IP) and out-of-plane (OOP) geometries. 
Going from low temperature the ZFC and FC curves start to overlap at approximately $180$~K defining the irreversibility temperature $T_{\mathrm{irr}}$.  
Moreover, at the same temperature both out-of-plane and in-plane FC curves change their character from typical for ferromagnet to paramagnetic-like. 
Since the Curie temperatures for pure Co and for Co-Pd system are far above room temperature, the presence of the irreversibility temperature is the evidence for superparamagnetic behavior and indicates that above this point the Co-Pd FM clusters are fully unblocked. 
Below the Curie temperature a~superparamagnetic particle has its own superspin which is a~sum of the individual magnetic moments of the atoms within the particle. 
According to the relation $V'E_{\mathrm{a}}' \propto k_{\mathrm{B}} T_{\mathrm{crit}}'$ every superparamagnetic cluster with volume $V'$ and anisotropy energy $E_{\mathrm{a}}'$ has its own blocking temperature $T_{\mathrm{crit}}'$ for the superspin. 
Therefore the gradual decrease of temperature below $T_{\mathrm{irr}}$ causes blocking of the smaller particles. 
In both IP and OOP ZFC curves the maximum visible at approximately $140$~K determines the average blocking temperature for the superspins of the blocked ferromagnetic clusters. 
This maximum is relatively broad with full width at half maximum of approximately $95$~K suggesting a~wide FM cluster size distribution.
This distribution is caused by the interface jaggedness in the multilayer which promotes the formation of Co-Pd FM clusters of various sizes and orientations.

The FC curve for in-plane geometry has a~plateau below the average FM blocking temperature. 
Such FC curve shape is observed for (super)spin glasses and superferromagnetic materials and indicates a~collective spin behavior with inter-particle interactions.\cite{Bed09}  
To determine the type of the interaction the magnetic relaxation measurements were carried out, and the results are presented in~\ref{Fig_2}b.   
The relaxation curves for both IP and OOP geometries (\ref{Fig_2}b) show a~magnetization decay with time following a~power-law $M(t) = M_{\infty} + M_{1} t^{\alpha}$,\cite{Bed09,Che03} where $M_{\infty}$ and $M_{1}$ are fit constants. 
The fitted values of the exponent $\alpha$ are $1.1$ for OOP and $1.05$ for IP geometries.  
The character of the curves as well as the values of the exponents indicates that the interactions between blocked particles have a~superferromagnetic nature. 
The observed magnetization decay is due to the domain wall motion through the blocked particles assembly. 
Additionally, the magnetization decay is faster for the IP geometry and demonstrates a~more easy move of the domain walls along this direction.    
Moreover, the relaxation curve for OOP geometry has a~minimum for $t \approx 1.5 \cdot 10^{5}$~s, characteristic for the superferromagnetic behavior.\cite{Che03}  
In such case the increase of the magnetization can be associated with superspin alignment within the magnetic domains. 
It was shown\cite{Che03} that this part can be described by adding a~saturating stretched exponential contribution $M_{2}\left( 1 - \exp\left[ -\left( t/\tau \right)^{\beta} \right] \right)$ ($\tau$ and $\beta$ are relaxation time and stretching exponent, respectively) to the power-law. 
The fit of this equation to the data fails, which can be explained by the interactions occurring not only between blocked FM particles but also between FM and AFM materials. 

\ref{Fig_2}c shows the out-of-plane hysteresis loops at 300~K for $[$Co/Pd$]_{10}$ and $[$CoO/Co/Pd$]_{10}$ multilayers. 
In both cases the thicknesses of Co and Pd are the same, and equal to 0.3~nm and 0.9~nm, respectively. 
The $[$Co/Pd$]_{10}$ system reveals a~clear ferromagnetism with out-of-plane easy axis due to the strong surface anisotropy of the Co/Pd interfaces.\cite{Car85, Car03} 
The introduction of the CoO, placed between the consecutive Co-Pd layers, leads to the superparamagnetic behavior at room temperature with no remanence and coercivity.  

\begin{figure*}[t]
\centering
\includegraphics[width=0.65\textwidth]{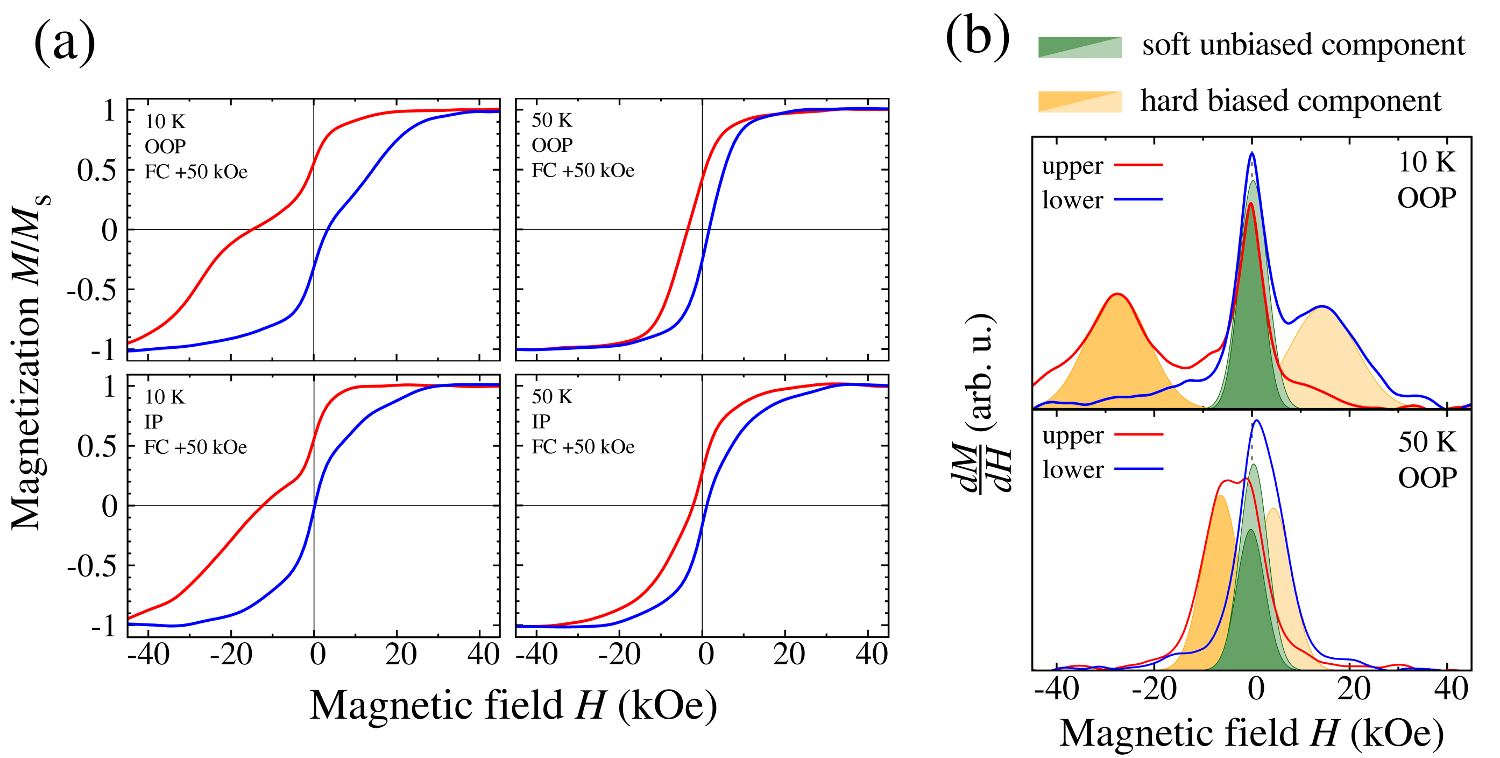}
\caption{(a) Magnetic hysteresis loops for $\lbrack$CoO/Co/Pd$\rbrack_{10}$ multilayer measured at $10$~K and 50~K after cooling in $+50$~kOe for out-of-plane and in-plane geometries. (b) First derivatives of the upper and lower branches of the OOP magnetization curves measured at 10~K and 50~K. Two main switching field components are indicated with colored areas.}
\label{Fig_3}
\end{figure*}

The information about the exchange anisotropy field and coercivity, together with their dependence on temperature, was obtained from the hysteresis loops. 
The measurements in both IP and OOP geometries were carried out after cooling from $300$~K down to desired temperature in external magnetic field of $+50$~kOe. 
The loops were measured in the field ranging from $+50$~kOe to $-70$~kOe, and both of these fields were large enough to saturate the system. 
Representative measurements at $10$~K and $50$~K for IP and OOP geometries are shown in~\ref{Fig_3}a.
A~negative loop shift from zero position is the evidence that the exchange anisotropy was induced in the system. 
At temperature $10$~K the observed exchange anisotropy field $H_{\mathrm{ex}}$ for both OOP and IP geometries is $6$~kOe. 
This loop shift is large in comparison to the other studies on CoO/Co system where the exchange anisotropy field of a~few hundred Oe was reported \cite{Dob12,Gru00,Dia14,Gie02,Gir03}. 
However, these studies were performed on systems with FM thickness of a few nm while in our case the FM layer is replaced by the Co-Pd FM clusters. 

We observe that the loops have asymmetric shape, which is especially pronounced at 10~K and suggests a~stepwise magnetic reversal process. 
The switching field distributions, calculated as first derivatives $dM$/$dH$ of the lower and upper branches of the 
OOP magnetization curves measured at 10~K and 50~K, are shown in~\ref{Fig_3}b. 
The two maxima observed in the $dM$/$dH$ curves can be associated with two major magnetic phases reversing at different external magnetic fields. 
The positions of the soft component maxima (green areas in~\ref{Fig_3}b), demonstrated for lower external field are centered around zero.  
The maxima of the hard component (orange areas in~\ref{Fig_3}b) are observed for larger external fields with a~clear bias along field axis. 
Therefore, this component can be identified as arising from the reversal process of the blocked FM particles coupled to the AFM grains. 
The unbiased soft component can be attributed to the blocked FM particles which are not interacting with the AFM material. 
The superposition of these two contributions results in the asymmetric loop shape seen in our results. 
Qualitatively similar behavior is observed for hysteresis loops obtained in IP geometry.

\begin{figure}[th]
\centering
\includegraphics[width=0.5\textwidth]{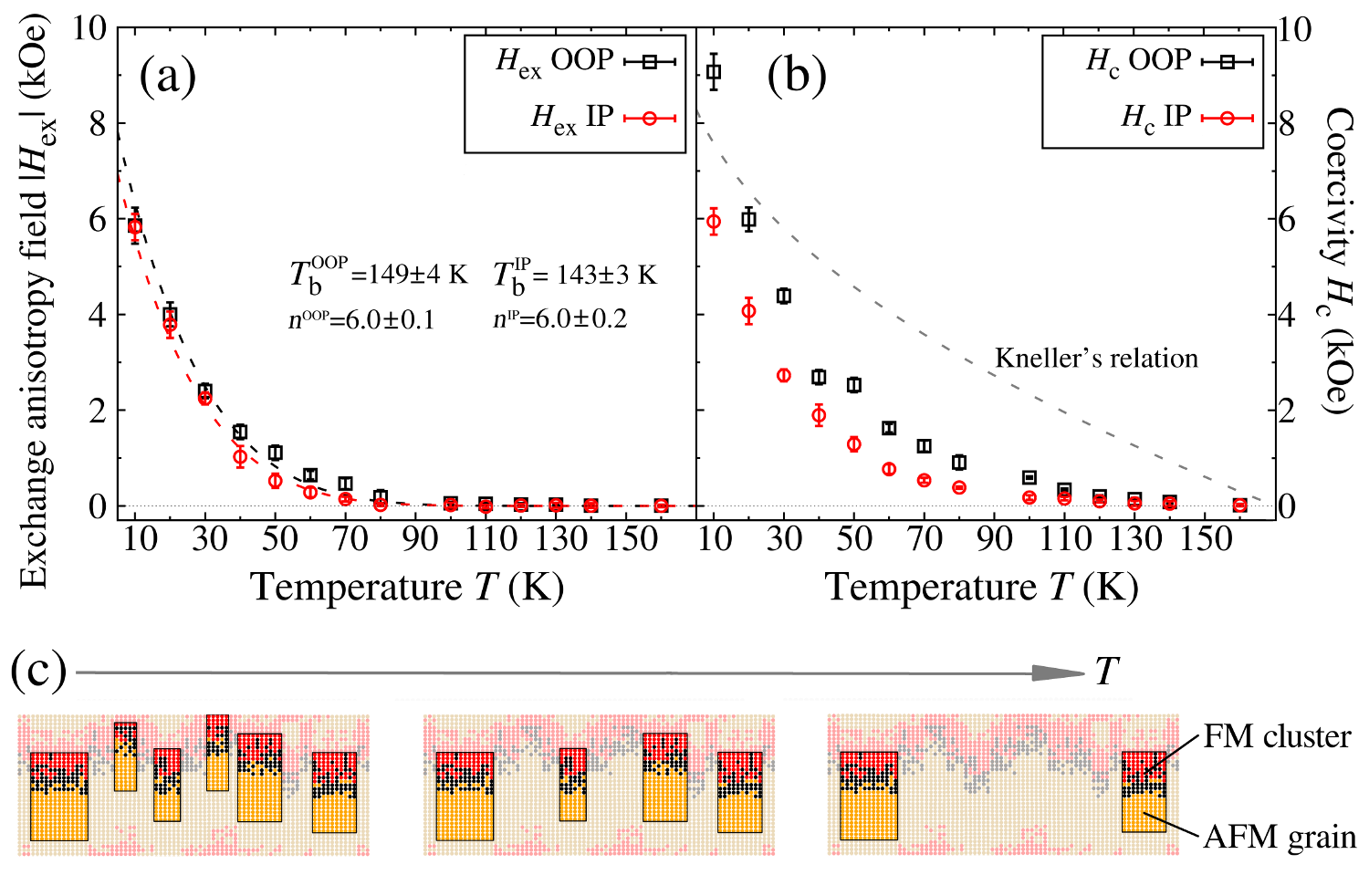}
\caption{Exchange anisotropy field $H_{\mathrm{ex}}$ (a) and coercivity $H_{\mathrm{c}}$ (b) dependencies on temperature $T$ for $\lbrack$CoO/Co/Pd$\rbrack_{10}$ multilayer measured for in-plane (IP) and out-of-plane (OOP) geometries. Dashed lines in Fig. (a) are fits. Line in Fig. (b) represents Kneller's relation for temperature dependence of coercivity in case of single-domain and non-interacting FM particles. Fig. (c) schematically presents the temperature evolution of the number of coupled FM/AFM grains.}
\label{Fig_4}
\end{figure}

The dependence of the exchange anisotropy field $H_{\mathrm{ex}}$ on temperature $T$ can be described by the relation $H_{\mathrm{ex}}\left(T\right) \propto \left(1-T/T_{\mathrm{b}}\right)^{n}$, where $T_{\mathrm{b}}$ is the blocking temperature for exchange bias and $n$ is a characteristic exponent \cite{Mal88}. 
Using this relation to fit both OOP and IP $H_{\mathrm{ex}}\left(T\right)$ dependencies (\ref{Fig_4}) we determined the average blocking temperature for exchange bias to be $T_{\mathrm{b}}\!=\!146$~K, which is much lower than Neel temperature for bulk CoO ($291$~K). 
The larger blocking temperatures ($175$~K -- $293$~K) were reported by others~\cite{Dob12, Gru00, Lam13, Kap03} for thicker CoO AFM layers. 
However, Zaag et al. \cite{Zaa00} found that reduction of the CoO thickness below $10$~nm results in lowering the exchange bias blocking temperature below bulk N\'eel temperature. 
According to their work, for CoO thickness of $1.8$~nm the blocking temperature for exchange bias is approx. $150$~K, which remains in agreement with our experimental observation from XRR and TEM measurements. 
The $H_{\mathrm{ex}}\left(T\right)$ dependencies for both OOP and IP directions show that the $H_{\mathrm{ex}}$ values are nearly the same. 
This means that during field cooling below blocking temperature $T_{\mathrm{b}}$ the same amount of AFM material is coupled to the blocked Co-Pd FM clusters.

By fitting the temperature dependence on the exchange anisotropy field $H_{\mathrm{ex}}\left(T\right)$ we determined the exponent~$n$. 
According to Malozemoff \cite{Mal88} this exponent is $n\!=\!1$ for cubic anisotropy of the AFM, and $n\!=\!1/2$ for its uniaxial anisotropy. 
Typically, in case of exchange biased systems based on CoO antiferromagnet, either a linear dependence is observed, or the exponent is slightly lower than unity.~\cite{Zho04,Dob12,Gru00,Lam13,Kap03,Men13} 
The exponent of $n\!=\!1.2$ has also been reported and explained with the thermal instabilities in the AFM order and oxide grain separation~\cite{Sah12}. 
In our case for both OOP and IP geometries the fitted exponent is $n\!=\!6$, which is a~result that has never been shown before for a CoO/Co-based system. 
This result suggests that a mechanism different than AFM symmetry is responsible for the fast rise of the exchange anisotropy field with decreasing temperature. 
We can see from the FC curves (\ref{Fig_2}a) that at temperature below $180$~K the superparamagnetic Co-Pd clusters become blocked and the system starts to reveal ferromagnetic behavior.  
Below the blocking temperature for the exchange bias $T_{\mathrm{b}}\!=\!146$~K the blocked FM regions couple to the antiferromagnetic grains resulting in the exchange anisotropy. 
Sahoo et al.~\cite{Sah12} suggested that coupling the FM layer to the independent AFM grains results in small increase of the exponent~$n$. 
However, the AFM domain separation alone cannot explain as large increase of the exponent~$n$ as observed in the results. 
The broad maximum observed in the ZFC curves (\ref{Fig_2}a) indicates that FM clusters with a~wide size distribution are present in the system. 
Therefore, the occurrence of the exchange coupling between single FM and AFM grains at certain temperature depends on whether the Co-Pd cluster is in a~blocked state. 
If the volume and anisotropy energy of the cluster are large enough to stabilize the superspin direction, then the superspin has an ability to be exchange coupled to the AFM grain (\ref{Fig_4}c, right side) contributing to the overall exchange anisotropy energy. 
In case when the FM cluster is thermally unstable, it does not couple to the nearby AFM grain. 
For lower temperature smaller Co-Pd FM clusters are blocked and coupled to the AFM domains and a larger number of FM/AFM grain pairs contribute to the exchange anisotropy energy (\ref{Fig_4}c, center and left side). 
Simultaneously, for each exchange coupled FM/AFM region lowering of the temperature makes the resulting exchange anisotropy field larger.
Therefore, the overall dependence of the exchange anisotropy field on temperature $H_{\mathrm{ex}}\left(T\right)$ rises from the contributions of the FM clusters, which become blocked at different temperatures. 
This mechanism is responsible for a~rapid rise of the exchange anisotropy field with decreasing temperature, and its influence dominates the contribution dependent on the anisotropy symmetry of the AFM material.

\begin{figure*}[t]
\centering
\includegraphics[width=0.75\textwidth]{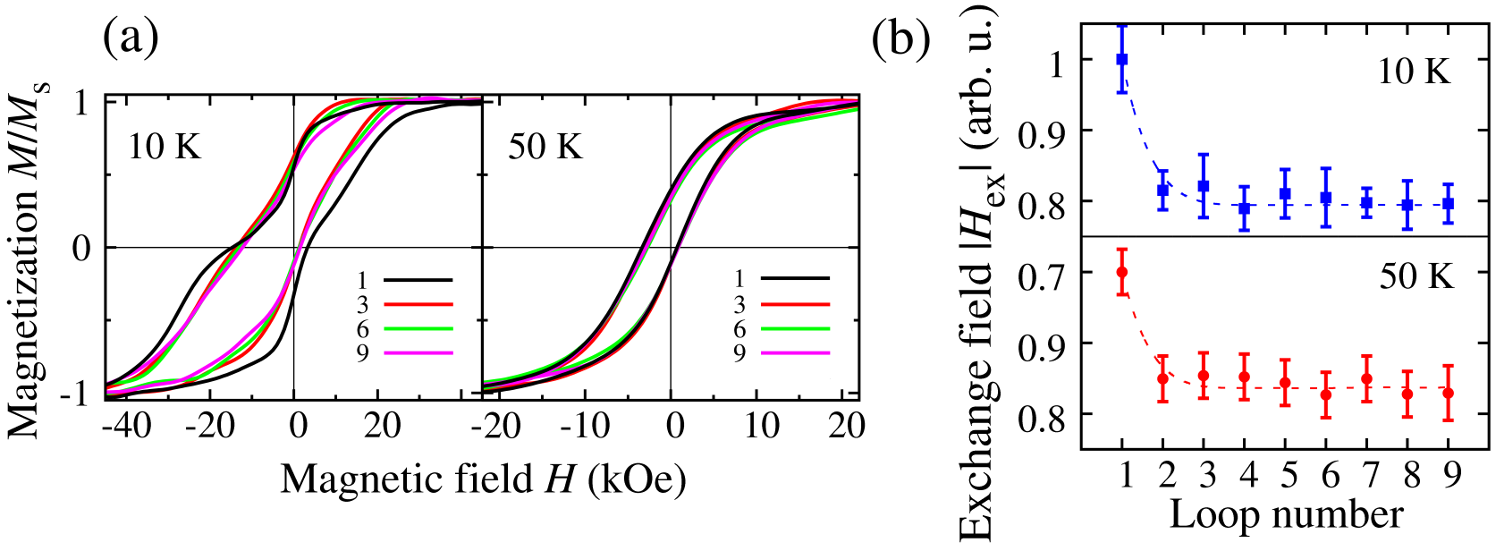}
\caption{(a) Training effect hysteresis loops measured in OOP geometry for $\lbrack$CoO/Co/Pd$\rbrack_{10}$ system at 10~K and 50~K. (b) The dependencies of the exchange anisotropy field $H_{\mathrm{ex}}$ on the loop number at 10~K and 50~K. The $H_{\mathrm{ex}}$ values were normalized to the exchange field measured for the first loop. The dotted lines are guides for the eye.}
\label{Fig_5}
\end{figure*}

Polycrystalline exchange-biased systems usually exhibit a~training effect, which manifests itself as a~monotonic decrease of the exchange anisotropy field through consecutive hysteresis loop cycling.
The presence of this effect is related to the spin structure rearrangement in the AFM material leading to the equlibrium configuration.~\cite{Bin04}
The training loops for the $\lbrack$CoO/Co/Pd$\rbrack_{10}$ system measured at 10~K and 50~K in OOP geometry are presented in \ref{Fig_5}a, and the dependencies of the exchange anisotropy field on the loop number are shown in \ref{Fig_5}b.
At both temperatures the bias field decreases only after first loop cycles and then stabilizes at a~constant level.
At temperature of 10~K the exchange bias field is reduced to the $0.8$ of the value measured for the first loop while at 50~K this reduction is slightly lower and equals $0.83$.  
The dependence of the exchange field on the loop number $H_{\mathrm{ex}}(n)$ does not follow the often observed power-law $H_{\mathrm{ex}}(n) - H_{\mathrm{ex}}(\infty) \propto 1/\sqrt{n}$, where $H_{\mathrm{ex}}(\infty)$ is the bias field in the limit of infinite loops. 
Moreover, the analytical model for training effect presented by Sahoo et al.~\cite{Sah07} and~applied by Wu et al.~\cite{Wu15} also does not fit to the data obtained for the investigated system. 
Our results differ from reported by others for CoO/Co systems\cite{Gru00,Dia14,Wu15,Ali12} where a~larger decrease of $H_{\mathrm{ex}}$ was shown together with its slower reduction with the cycle number. 
A~comparable training reduction of the exchange bias field for CoO/Co system was shown by Biniek et al.\cite{Bin05}. 
However, in this case the decrease of $H_{\mathrm{ex}}$ was also more gradual than the sharp one-step drop observed in our results. 
We want to emphasize that these previous studies looked at systems with continuous AFM and FM layers and their results can be explained using the domain state model.\cite{Now02}
In our case the exchange coupling does not take place all over the continuous interface between FM and AFM materials and can be addressed as local interactions between blocked FM clusters and AFM grains (see~\ref{Fig_4}c). 
Therefore, the AFM spin rearrangement induced by the FM reversal process is limited to the smaller volume than the whole AFM layer which may restrict the possible AFM domain wall motion and AFM domain reorientation processes.   
Because of this, after the second field cycle the spin structures of the AFM grains freeze in the~locally favorable spin configurations and prevents further spin relaxation through the wall motion.

The temperature dependencies of coercivity $H_{\mathrm{c}}\left(T\right)$ (\ref{Fig_4}b), observed for both OOP and IP directions, show similar exponential growth for decreasing temperature as $H_{\mathrm{ex}}\left(T\right)$ data (\ref{Fig_4}a).
One can expect that the coercivity of the assembly of small particles will follow the Kneller's temperature dependence $H_{\mathrm{c}}\left(T\right) \propto [ 1-(T/T_{\mathrm{crit}})^{1/2} ]$~\cite{Ali14}, where $T_{\mathrm{crit}}$ is a~temperature above which FM clusters are fully unblocked.
In our case the following relation is not fulfilled which reveals that the magnetization reversal process present in our system cannot be described by the Stoner-Wohlfarth model.
The reasons for this are twofold.  
First, the blocked FM clusters show a~superferromagnetic collective behavior and cannot be treated as non-interacting objects. 
Second, there is a~permanent growth of the number of blocked FM clusters with decreasing temperature.    
Since the new clusters become blocked, they start to contribute to the overall coercivity of the system, and the situation is similar to the case of the exchange anisotropy field, which was discussed earlier.

\begin{figure}[h]
\centering
\includegraphics[width=0.47\textwidth]{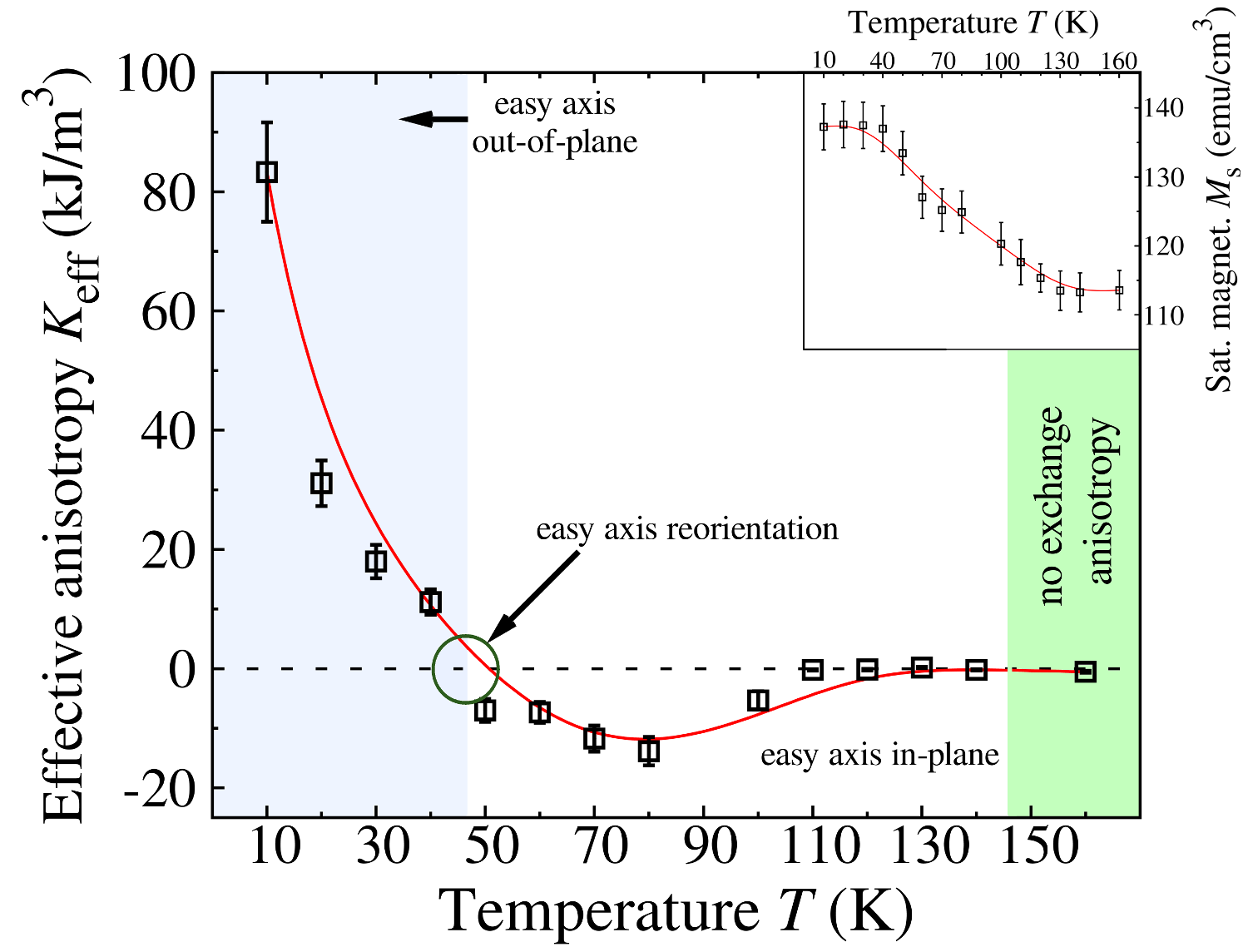}
\caption{Temperature dependence of the effective anisotropy energy $K_{\mathrm{eff}}\left(T\right)$, obtained from hysteresis loops measured for field-cooled $\lbrack$CoO/Co/Pd$\rbrack_{10}$ multilayer. The inset shows the temperature dependence of saturation magnetization $M_{\mathrm{s}}\left(T\right)$. The red lines are guides for the eye.}
\label{Fig_6}
\end{figure}

The area under a~hysteresis loop is a~measure of the magnetic energy stored in the system after cooling in certain conditions. 
The half of a~difference between areas under OOP and IP loops is called effective anisotropy energy $K_{\mathrm{eff}}$ and its sign is dependent on the preferred easy axis of magnetization orientation. 
In our case it is positive for the out-of-plane configuration and negative for the in-plane case. 
The determined temperature dependence $K_{\mathrm{eff}}(T)$ is shown in~\ref{Fig_6}. 
Below the blocking temperature for exchange bias $T_{\mathrm{b}}\!=\!146$~K the effective anisotropy energy is negative and decreases down to the minimum at approx. $80$~K. 
In this temperature range the main contribution to the energy $K_{\mathrm{eff}}$ is the demagnetization energy, which is proportional to the second power of the saturation magnetization and favors the in-plane spin configuration. 
Since the saturation magnetization of the system slowly increases with decreasing temperature (inset in \ref{Fig_6}) the in-plane spin orientation becomes more favorable. 
Effective anisotropy energy $K_{\mathrm{eff}}$ starts to grow below 80~K. 
Since the Co-Pd clusters have large out-of-plane anisotropy energy\cite{Car85,Car03} the progressive blocking of such objects introduces an increasing out-of-plane magnetization component as more clusters become blocked. 
The constant increase of the blocked FM clusters number for the lower temperature results in further enhancement of perpendicular anisotropy energy and causes an easy axis switching at approximately 50~K when the out-of-plane magnetization component becomes dominant over the demagnetization energy.   
The effect of the easy axis switching with temperature was reported for the exchange biased permalloy/CoO multilayer by Zhou et al.\cite{Zho04}. 
However, in the case described in that study the FM layers were continuous and the reorientation process was driven only by the surface anisotropy of the interfaces, without any influence of superparamagnetism. 
In our case the observed increase of the perpendicular surface anisotropy from Co/Pd interface below $50$~K is connected with the temperature distribution of the number of blocked ferromagnetic clusters.

\section{Conclusions}

In this paper we have shown the influence of the superparamagnetism on exchange anisotropy and the orientation of the magnetization easy axis in the case of $\lbrack$CoO/Co/Pd$\rbrack_{10}$ system. 
We have found that the decrease of the Co thickness below the limit for continuous layer formation leads to the creation of the ferromagnetic Co-Pd clusters placed between consecutive AFM CoO grains. 
The FM particles interact with each other and show superferromagnetic collective behavior. 
In such case the exchange coupling between FM particles and the AFM material results in exchange anisotropy field out to 6~kOe. 
The Co-Pd FM clusters start to block their superspins below $180$~K, and below the blocking temperature for exchange bias, which is 146~K, they couple to the CoO antiferromagnetic grains. 
The observed unusual rapid rise of the exchange anisotropy field with decreasing temperature is connected with the gradual process of thermal blocking of the FM clusters superspin. 
The increased number of the blocked FM particles gives rise to the coercivity resulting in its fast enhancement for lower temperature. 
Since the Co-Pd clusters have a~large out-of-plane surface anisotropy the process of the FM cluster thermal blocking affects also the orientation of the easy magnetization axis and causes the axis switching to out-of-plane direction at approximately~50~K.

\bibliographystyle{plainnat}

\begin{thebibliography}{10}

\bibitem{Kiw01} Kiwi, M. Exchange Bias Theory. \textit{J. Magn. Magn. Mater.} \textbf{2001}, \textit{234}, 584 -- 595.

\bibitem{Nog99} Nogu\'es, J.; Schuller, I. K. Exchange Bias. \textit{J. Magn. Magn. Mater.} \textbf{1999}, \textit{192},  203 -- 232.

\bibitem{Nog05} Nogu\'es, J.; Sort, J.; Langlais, V.; Skumryev, V.; Suri\~{n}ach, S.; Mu\~{n}oz, J. S.; Bar\'o, M. D. Exchange Bias in Nanostructures. \textit{Phys. Rep.} \textbf{2005}, \textit{422}, 65 -- 117.

\bibitem{Zho04} Zhou, S. M.; Sun, L.; Searson, P. C.; Chien, C. L. Perpendicular Exchange Bias and Magnetic Anisotropy in CoO/Permalloy Multilayers. \textit{Phys. Rev. B} \textbf{2004}, \textit{69}, 024408. 

\bibitem{Sta00} Stamps, R. L. Mechanisms for Exchange Bias. \textit{J. Phys. D: Appl. Phys.} \textbf{2000}, \textit{33}, R247 -- R268. 

\bibitem{Pol14} Polenciuc, I.; Vick, A. J.; Allwood, D. A.; Hayward, T. J.; Vallejo-Fernandez, G.; O'Grady, K.; Hirohata, A. Domain Wall Pinning for Racetrack Memory Using Exchange Bias. \textit{Appl. Phys. Lett.} \textbf{2014}, \textit{105}, 162406. 

\bibitem{Kle07} Klem, M. T.; Resnick, D. A.; Gilmore, K.; Young, M.; Idzerda, Y. U.; Douglas, T. Synthetic Control over Magnetic Moment and Exchange Bias in All-Oxide Materials Encapsulated within a~Spherical Protein Cage. \textit{J. Am. Chem. Soc.} \textbf{2007}, \textit{129}, 197 -- 201.

\bibitem{Iss13} Issa, B.; Obaidat, I. M..; Albiss, B. A.; Haik, Y. Magnetic Nanoparticles: Surface Effects and Properties Related to Biomedicine Applications. \textit{Int. J. Mol. Sci.} \textbf{2013}, \textit{14}, 21266 -- 21305.

\bibitem{Ber99} Berkowitz, A. E.; Takano, K. Exchange Anisotropy --- A Review. \textit{J. Magn. Magn. Mater.} \textbf{1999}, \textit{200}, 552 -- 570. 

\bibitem{Men14} Men\'endez, E.; Modarresi, H.; Dias, T.; Geshev, J.; Pereira, L. M. C.; Temst, K.; Vantomme, A. Tuning the Ferromagnetic-Antiferromagnetic Interfaces of Granular Co-CoO Exchange Bias Systems by Annealing. \textit{J. Appl. Phys.} \textbf{2014}, \textit{115}, 133915. 

\bibitem{Dob12} Dobrynin, A. N.; Givord, D. Exchange Bias in a~Co/CoO/Co Trilayer with Two Different Ferromagnetic-Antiferromagnetic Interfaces. \textit{Phys. Rev. B} \textbf{2012}, \textit{85}, 014413.

\bibitem{Gru00} Gruyters, M.; Riegel, D.; Strong Exchange Bias by a~Single Layer of Independent Antiferromagnetic Grains: The CoO/Co Model System. \textit{Phys. Rev. B} \textbf{2000}, \textit{63}, 052401.

\bibitem{Car85} Carcia, P. F.; Meinhaldt, A. D.; Suna, A. Perpendicular Magnetic Anisotropy in Pd/Co Thin Film Layered Structures. \textit{Appl. Phys. Lett.} \textbf{1985}, \textit{47}, 178 -- 180.

\bibitem{Car03} Carrey, J.; Berkowitz, A. E.; Egelhoff, W. F.; Smith, D. J. Influence of Interface Alloying on the Magnetic Properties of Co/Pd Multilayers. \textit{Appl. Phys. Lett.} \textbf{2003}, \textit{83}, 5259 -- 5261.

\bibitem{Raf02} Rafaja, D.; Fuess, H.; \v{S}imek, D.; Kub, J.; Zweck, J.; Vac\'{i}nov\'{a}, J.; Valvoda, V. X-Ray Reflectivity of Multilayers with Non-continuous Interfaces. \textit{J. Phys.: Condens. Matter} \textbf{2002}, \textit{14}, 5303 -- 5314. 

\bibitem{Bed09} Bedanta, S.; Kleemann, W. Supermagnetism. \textit{J. Phys. D: Appl. Phys.} \textbf{2009}, \textit{42}, 013001.

\bibitem{Che03} Chen, Xi; Kleemann, W.; Petracic, O.; Sichelschmidt, O.; Cardoso, S.; Freitas, P. Relaxation and Aging of a~Superferromagnetic Domain States. \textit{Phys. Rev. B} \textbf{2003}, \textit{68}, 054433.

\bibitem{Dia14} Dias, T.; Men\'endez, E.; Liu, H.; Van Haesendonck, C.; Vantomme, A.; Temst, K.; Schmidt, J. E.; Giulian, R.; Geshev, J. Rotatable Anisotropy Driven Training Effects in Exchange Biased Co/CoO Films \textit{J. Appl. Phys.} \textbf{2014}, \textit{115}, 243903. 

\bibitem{Gie02} Gierlings, M.; Prandolini, M. J.; Fritzsche, H.; Gruyters, M.; Riegel, D. Change and Asymmetry of Magnetization Reversal for a Co/CoO Exchange-Bias System \textit{Phys. Rev. B} \textbf{2002}, \textit{65}, 092407.

\bibitem{Gir03} Girgis, E.; Portugal, R. D.; Loosvelt, H.; Van Bael, M. J.; Gordon, I.; Malfait, M.; Temst, K.; Van Haesendonck, C.; Leunissen, L. H. A.; Jonckheere, R. Enhanced Asymmetric Magnetization Reversal in Nanoscale Co/CoO Arrays: Competition Between Exchange Bias and Magnetostatic Coupling \textit{Phys. Rev. Lett.} \textbf{2003}, \textit{91}, 187202. 

\bibitem{Mal88} Malozemoff, A. P. Mechanisms of Exchange Anisotropy. \textit{J. Appl. Phys.} \textbf{1988}, \textit{63}, 3874 -- 3879. 

\bibitem{Lam13} Lamirand, A. D.; Ramos, A. Y.; De Santis, M.; Cezar, J. C.; De Siervo, A.; Jamet, M. Robust Perpendicular Exchange Coupling in an Ultrathin CoO/PtFe Double Layer: Strain and Spin Orientation. \textit{Phys. Rev. B} \textbf{2013}, \textit{88}, 140401(R).

\bibitem{Kap03} Kappenberger, P.; Martin, S.; Pellmont, Y.; Hug, H. J.; Kortright, J. B.; Hellwig, O.; Fullerton, E. E. Direct Imaging and Determination of the Uncompensated Spin Density in Exchange-Biased CoO/(CoPt) Multilayers. \textit{Phys. Rev. Lett.} \textbf{2003}, \textit{91}, 267202. 

\bibitem{Zaa00} Van Der Zaag, P. J.; Ijiri, Y.; Borchers, J. A.;. Feiner, L. F.; Wolf, R. M.; Gaines, J. M.; Erwin, R. W.; Verheijen, M. A. Difference Between Blocking and N\'eel Temperatures in the Exchange Biased Fe$_{3}$O$_{4}$/CoO System. \textit{Phys. Rev. Lett.} \textbf{2000}, \textit{84}, 6102 -- 6105. 

\bibitem{Men13} Men\'endez, E.; Demeter, J.; Van Eyken, J.; Nawrocki, P.; Jedryka, E.; W\'ojcik, M.; Lopez-Barbera, J. F.; Nogu\'es, J.; Vantomme, A.; Temst, K. Improving the Magnetic Properties of Co-CoO Systems by Designed Oxygen Implantation Profiles. \textit{ACS Appl. Mater. Interfaces} \textbf{2013}, \textit{5}, 4320 -- 4327. 

\bibitem{Sah12} Sahoo, S.; Polisetty, S.; Wang, Y.; Mukherjee, T.; He, X.; Jaswal, S. S.; Binek, C. Asymmetric Magnetoresistance in an Exchange Bias Co/CoO Bilayer. \textit{J. Phys.: Condens. Matter} \textbf{2012}, \textit{24}, 096002. 

\bibitem{Bin04} Binek, C. Training of the Exchange-Bias Effect: A~Simple Analytic Approach. \textit{Phys. Rev. B} \textbf{2004}, \textit{70}, 014421.

\bibitem{Sah07} Sahoo, S.; Polisetty, S.; Binek, C.; Berger, A. Dynamic Enhancement of the Exchange Bias Training Effect. \textit{J. Appl. Phys.} \textbf{2007}, \textit{101}, 053902.

\bibitem{Wu15} Wu, R.; Fu, J. B.; Zhou, D.; Ding, S. L.; Wei, J. Z.; Zhang, Y.; Du, H. L.; Wang, C. S.; Yang, Y. C.; Yang, J. B. Temperature Dependence of Exchange Bias and Training Effect in Co/CoO Film with Induced Uniaxial Anisotropy. \textit{J. Phys. D: Appl. Phys.} \textbf{2015}, \textit{48}, 275002.

\bibitem{Ali12} Ali, S. R.; Ghadimi, M. R.; Fecioru-Morariu, M.; Beschoten, B.; G\"untherodt, G. Training Effect of the Exchange Bias in Co/CoO Bilayers Originates from the Irreversible Thermoremanent Magnetization of the Magnetically Diluted Antiferromagnet. \textit{Phys. Rev. B} \textbf{2012}, \textit{85}, 012404. 

\bibitem{Bin05} Binek, C.; He, X.; Polisetty, S. Temperature Dependence of the Training Effect in a~Co/CoO Exchange-Bias Layer. \textit{Phys. Rev. B} \textbf{2005}, \textit{72}, 054408.

\bibitem{Now02} Nowak,~U.; Usadel, K. D.; Keller, J.; Milt\'{e}nyi, P.; Beschoten, B.; G\"untherodt, G. Domain State Model for Exchange Bias. I. Theory. \textit{Phys. Rev. B} \textbf{2002}, \textit{66}, 014430.


\bibitem{Ali14} Ali, K.; Sarfraz, A. K.; Ali, A.; Mumtaz, A.; Hasanain, S. K. Temperature Dependence Magnetic Properties and Exchange Bias Effect in CuFe$_{2}$O$_{4}$ Nanoparticles Embedded in NiO Matrix. \textit{J. Magn. Magn. Mater.} \textbf{2014}, \textit{369}, 81 -- 85. 


\end{thebibliography}
\vspace{1.1ex}
\begin{center}
 $\star$ $\star$ $\star$
\end{center}

\vspace{-9ex}

% styl listowania wpisów bibiograficznych
\setlength{\bibsep}{0pt}
\renewcommand{\bibnumfmt}[1]{$^{#1}$}

\end{document}